\documentclass[12pt]{iopart}

\usepackage{graphicx}
\usepackage{mathptmx, courier, pifont}
\usepackage[scaled=0.92]{helvet}
\usepackage[T1]{fontenc}
\usepackage{textcomp}

\newlength{\figdn}
\newcommand{\fplus}[1]{F^+_{#1}}
\newcommand{\fminus}[1]{F^-_{#1}}
\newcommand{\eq}[1]{Eq.~(\ref{#1})}
\newcommand{\eqnoeq}[1]{(\ref{#1})}
\newcommand{\fig}[1]{Fig.~\ref{fig:#1}}
\newcommand{\deriv}[2]{\frac{d#1}{d#2}}
\newcommand{\tsub}[1]{_{\mbox{\scriptsize#1}}}
\newcommand{\tsup}[1]{^{\mbox{\scriptsize#1}}}
\newcommand{\tfrac}[2]{\mbox{$\small\frac{#1}{#2}$}}
\newcommand{\units}[1]{\mbox{\ #1}}
\newcommand{\isotope}[2]{\mbox{$^{#1}$#2}}

\newcommand{\singlefig}[6]{%
\begin{figure} \vspace{#3}%
\begin{flushright}%
\includegraphics*[scale=#5]{#2}%
\end{flushright}%
\caption{\label{fig:#1} #6}%
\vspace{#4}%
\end{figure}}

\newcommand{\putfig}[6]{%
\begin{figure}\vspace*{#2}%
\begin{flushright}%
\includegraphics*[scale=#4]{#5}%
\end{flushright}%
\caption{#6}%
\vspace*{#3}%
\label{fig:#1}%
\end{figure}}

\begin{document}

\title
{Explicit Integration of Extremely-Stiff Reaction
Networks: Quasi-Steady-State Methods}

\author{M. W. Guidry$^{1,2,3}$ and J. A. Harris$^1$}

\address{$^1$Department of Physics and Astronomy, University of Tennessee,
Knoxville, TN 37996-1200, USA}
\address{$^2$Physics Division, Oak Ridge National Laboratory, Oak Ridge, TN
37830, USA}
\address{$^3$Computer Science and Mathematics Division, Oak Ridge National
Laboratory, Oak Ridge, TN 37830, USA}

\ead{guidry@utk.edu}

\begin{abstract}
A preceding paper  \cite{guidAsy} demonstrated that explicit asymptotic methods
generally work much better for extremely stiff reaction networks than has
previously been shown in the literature. There we showed that for systems well
removed from equilibrium explicit asymptotic methods can rival standard implicit
codes in speed and accuracy for solving extremely stiff differential equations.
In this paper we continue the investigation of systems well removed from
equilibrium by examining quasi-steady-state (QSS) methods as an alternative to
asymptotic methods. We show that for systems well removed from equilibrium, QSS
methods also can compete with, or even exceed, standard implicit methods in
speed, even for extremely stiff networks, and in many cases give somewhat better
integration speed than for asymptotic methods. As for asymptotic methods, we
will find that QSS methods give correct results, but with non-competitive
integration speed as equilibrium is approached. Thus, we shall find that both
asymptotic and QSS methods must be supplemented with partial equilibrium methods
as equilibrium is approached to remain competitive with implicit methods. 
\end{abstract}


% Keywords required only for MST, PB, PMB, PM, JOA, JOB? 
\vspace{2pc}

% Uncomment for Submitted to journal title message
%\submitto{\JPA}

% Comment out if separate title page not required
%\maketitle

\section{Introduction}

Stiff networks of differential equations rather uniformly have  been viewed as
requiring special implicit or semi-implicit methods for integration in order to
maintain stability while taking reasonably efficient timesteps
\cite{gear71,lamb91,press92,oran05,timmes,hix05}. Purely explicit methods are
not competitive in speed for most applications because they are limited by
stability criteria to integration timesteps that are far too short. Various
asymptotic and steady-state schemes have been proposed to stabilize explicit
methods by removing some of their stiffness (overviews may be found in Refs.\
\cite{oran05,mott99}). These methods have had some success in moderately stiff
systems, but it generally has been concluded that in very stiff systems, such as
those encountered in astrophysical thermonuclear networks, asymptotic and
steady-state schemes do not work \cite{oran05,mott99}. 

In a preceding paper on asymptotic methods \cite{guidAsy}, the present paper on
quasi-steady-state (QSS) methods, and a following paper on partial equilibrium
methods \cite{guidPE}, we challenge these conclusions and present strong
evidence that algebraically-stabilized explicit integration may in fact not only
compete with, but in some cases may have the potential to outperform
traditional implicit methods, even for the stiffest networks. In this paper we
deal specifically with the QSS method and show that, for systems well removed
from equilibrium, the QSS approximation can give highly-competitive integration
of extremely-stiff systems.

\section{\label{sh:steadystate} Quasi-Steady-State Approximations}

Let us begin by introducing the quasi-steady-state approximation.
We wish to solve  $N$ coupled ordinary differential equations 
\begin{eqnarray}
    \frac{dy_i}{dt} &=& F_i(y,t) = \sum_j F_{ij}(t) 
\nonumber
\\
&\equiv& \fplus i (t)- \fminus i(t)
    = \fplus i (t) - k_i(t) y_i(t)
\label{eq1.1}
\end{eqnarray}
where $y_i (i=1 \dots N)$ describes the dependent (abundance) variables, $t$ is
the independent variable (time in our examples), $F_{ij}$ denotes the flux
between species $i$ and $j$, the sum for each variable $i$ is over all variables
$j$ coupled to $i$ by a non-zero flux $F_{ij}$, and the flux has been decomposed
into a component $\fplus i$ increasing the abundance of $y_i$ and a component
$\fminus i$ depleting it. For an $N$-species network there will be $N$ such
equations in the populations $y_i$, and they generally will be coupled to each
other because of the dependence of the fluxes  on the different $y_j$.

If one attempts to integrate these equations numerically by ordinary forward
difference, severe stability problems will be encountered for networks in which
the various rate parameters appearing in the terms on the right side of
\eq{eq1.1} range over many orders of magnitude in size. This is the problem of
{\em stiffness}. The traditional solution is to invoke implicit methods, which
are stable even in the face of extremely stiff equations. An alternative
explicit algebraic solution to the coupled differential equations uses the
Quasi-Steady-State (QSS) approximations developed by Mott and
collaborators \cite{mott99,mott00}, which was partially motivated by earlier
work in Refs.\
\cite{verw94,verw95,jay97}.
We follow Mott et al
\cite{mott99,mott00} by first noting that \eq{eq1.1} in the form
\begin{equation}
 \deriv yt = F^+ (t) - k(t) y(t) \qquad y(0) \equiv y_0
\label{qss1.1a}
\end{equation}
(where we have suppressed indices for notational convenience) has the analytical
solution
\begin{equation}
 y(t) = y_0 e^{-kt} + \frac{F^+}{k}(1-e^{-kt}),
\label{qssSolution}
\end{equation}
for constant $k$ and $F^+$. In the QSS method this equation then serves
as the basis of a predictor--corrector scheme in which a prediction is made
using initial values and a corrector is then applied that uses a
combination of initial values and values computed using the predictor solution.
Defining a parameter $\alpha(r)$ by
\begin{equation}
\alpha(r) = \frac{160 r^3 + 60 r^2 + 11r +1}{360r^3 + 60r^2 + 12r +1},
\label{qss1.1}
\end{equation}
where $r \equiv 1/k\Delta t$ with $\Delta t$ the integration timestep, we adopt
a predictor $y\tsup p$ and corresponding corrector $y\tsup c$ proposed
originally by Mott et al \cite{mott99,mott00},
\begin{equation}
 y\tsup p = y^0 + \frac{\Delta t (F_0^+ - F_0^-)}{1 + \alpha^0 k^0 \Delta t}
\qquad
y\tsup c = y^0 + \frac{\tilde F^+ - \bar k y^0}{1+\bar\alpha \bar k \Delta t} ,
\label{qss1.2}
\end{equation}
where $\alpha^0$ is evaluated from \eq{qss1.1} with $r = 1/k^0 \Delta t$,
an average rate parameter is defined by 
$\bar k = \tfrac12 (k^0 + k\tsup p)$,
$\bar\alpha$ is specified by \eq{qss1.1} with $r = 1/\bar k\Delta t$, and
$$
 \tilde F^+ = \bar\alpha F_{\scriptstyle\rm p}^+ + (1-\bar\alpha)F_0^+.
%\label{qss1.5}
$$
If desired, the corrector can be iterated by using $y\tsup c$ from one iteration
step as the $y\tsup p$ for the next iteration step. We implement an explicit QSS
algorithm based on the predictor--corrector pair \eqnoeq{qss1.2} in a manner
analogous to that described in the preceding paper for the asymptotic method
\cite{guidAsy}, except that for the QSS algorithm we treat all equations by the
QSS approximation, rather than dividing them into a set treated by
explicit forward difference and a set treated in asymptotic approximation, as we
did in Ref.\ \cite{guidAsy}.

\section{Adaptive Timestepping}

To integrate the equations \eqnoeq{eq1.1} using the predictor--corrector
\eqnoeq{qss1.2}, we employ a simple timestepping algorithm analogous to that
already described in more detail in the preceding asymptotic paper
\cite{guidAsy}:

\begin{enumerate}
 \item 
At the beginning of a new timestep, compute a trial timestep based on limiting
the change in population that would result from that timestep to a specified
tolerance. Choose the minimum of this trial timestep and the timestep that was
taken in the previous integration timestep as the timestep, and update all
populations by the quasi-steady-state algorithm described above.
\item
Check the results for conservation of particle number within a specified
tolerance range.     If the condition is not satisfied, increase or decrease the
timestep as appropriate by a small factor and repeat the calculation with the
original fluxes.  Accept this result for the populations and carry the new
timestep over as a starting point for the next timestep.
\end{enumerate}
This timestepper is not particularly sophisticated but we have found it to be
stable and accurate for a variety of astrophysical thermonuclear networks that
we have tested.

\section{\label{sh:large-stiff} Equilibrium and Stiffness}

As we have discussed in more detail in Refs.\ \cite{guidAsy,guidPE}, there are
two forms of equilibrium that concern us in explicit integration of stiff
reaction networks.  These may be displayed clearly if we decompose $\fplus i$
and $\fminus i$ for a species $i$ in \eq{eq1.1} into a set of terms depending on
the other populations  in the network (labeled by the index $j$), 
\begin{eqnarray}
\deriv{y_i}{t} &=&
\fplus{i} - \fminus{i} 
\nonumber
\\
&=& (f_1^+ + f_2^+ + \ldots)_i -  (f_1^- + f_2^- + \ldots)_i
\nonumber
\\
&=& (f_1^+
- f_1^-)_i + (f_2^+ - f_2^-)_i + \ldots = \sum_j (f^+_j - f^-_j)_i,
\label{equilDecomposition}
\end{eqnarray}
We shall refer to {\em macroscopic equilibration} if $\fplus i - \fminus i$
approaches a constant.  This is the basis for the asymptotic approximations
discussed in Ref.\ \cite{guidAsy} and the quasi-steady-state approximation to be
discussed in this paper. However, at a more microscopic level, groups of
individual terms on the right side of \eq{equilDecomposition} may come
approximately into equilibrium (so that the sum of their fluxes is approximately
zero), even if the macroscopic conditions for equilibration are not satisfied.
This corresponds to equilibration for individual forward--reverse reaction pairs
such as $A+B+\ldots \rightleftharpoons C+D+\ldots$ .  This process, which may
occur even if the conditions for macroscopic equilibration are not satisfied, we
shall term {\em microscopic equilibration}. These definitions then permit us to
identify three distinct categories of stiffness that may occur in a reaction
network \cite{guidAsy}:

\begin{enumerate}
 \item 
Small populations can become negative if the explicit timestep is too large,
with the propagation of this anomalous negative population leading to
destabilizing terms that grow exponentially.
\item
Macroscopic equilibration, where taking the difference $\fplus{} - \fminus{}$
leads to large errors if the timestep is too large.
\item
Microscopic equilibration, where taking the net flux in specific forward-reverse
reaction pairs $(f^+_i - f^-_i)$ leads to large errors if the timestep is too
large. 
\end{enumerate}
These distinctions are crucial for our goal of integrating stiff equations
explicitly by identifying sources of stiffness in the network and removing them
by algebraic means because the QSS method and asymptotic methods  remove only
the first two kinds of stiffness. Removal of the third kind of stiffness will
require the partial equilibrium methods that will be discussed in the
third paper in this series \cite{guidPE}. Thus, it will be important for our
discussion to have a quantitative measure of microscopic equilibration. We shall
describe this in detail in Ref.\ \cite{guidPE} for thermonuclear networks, and
the basics have been worked out by Mott \cite{mott99}, so we just quote without
proof the results that will be relevant for the present discussion.

We assume that the amount of microscopic equilibration in a network is measured
by the fraction of forward--reverse reaction pairs  $A+B+\ldots
\rightleftharpoons C+D+\ldots$ that are judged to be in equilibrium (with each
reaction pair considered separately). The variation of the populations $y_i$
with time during a
numerical integration timestep may be approximated for each reaction pair by a
differential equation
\begin{equation}
 \deriv{y_i}{t} = a y_i^2 + b y_i + c,
\label{2body1.3}
\end{equation}
where the parameters $a$, $b$, and $c$ are known functions of the current rate
parameters and the abundances at the beginning of the timestep. This equation
may be solved for the equilibrium abundance $\bar y_i$ of each
species, giving
\begin{equation}
 \bar y_i \equiv y^{{\rm\scriptstyle eq}}_i = -\frac{1}{2a} (b + \sqrt{-q}).
\label{2body1.6}
\end{equation}
where $q \equiv 4ac-b^2$ and the single timescale $\tau = q^{-1/2}$
governs the approach to equilibrium. We may then estimate
whether a given reaction is near equilibrium at time $t$ by requiring
\begin{equation}
 \frac{| y_i(t) - \bar y_i |}{\bar y_i}
< \epsilon_i
\label{2body1.8b}
\end{equation}
for each species $i$ involved in the reaction, where $y_i(t)$ is the actual
abundance, $\bar y_i$ is the equilibrium abundance  \eqnoeq{2body1.6}, and
$\epsilon_i$ is a tolerance, typically of order $10^{-2}$. The remainder of this
paper will emphasize methods
based on quasi-steady-state approximations to stabilize explicit integration for
networks that are at most weakly equilibrated by the above criteria,
with the corresponding stabilization of networks near equilibrium to be
discussed in \cite{guidPE}.

\section{Explicit and Implicit Integration Speeds for a Timestep}

In examples to be shown below we shall be comparing explicit and implicit
methods using codes that are at very different stages of development and
optimization. Thus they cannot simply be compared head to head. Implicit methods
spend increasing amounts of integration time inverting matrices as networks
become larger. Thus, explicit methods---which require no matrix inversions---can
generally compute each timestep faster. Let us assume roughly that the speedup
factor for explicit versus implicit methods for integrating a timestep is
$F=1/(1-f)$, where $f$ is the fraction of computing time spent by the implicit
algorithm in matrix operations. Using data obtained by Feger
\cite{feg11a,feg11b} with the implicit, backward-Euler code Xnet \cite{raphcode}
employing both dense and sparse matrix solvers, we adopt for our discussion the
factors $F$ listed in Table \ref{tb:explicitSpeedup}. 
\begin{table}
\caption{\label{tb:explicitSpeedup}Explicit-method speedup factors
\cite{feg11b}}
\begin{indented}
\item[]\begin{tabular}{@{}lll}
\br
Network&Isotopes&Speedup $F$\\
\mr
CNO (main) & 8 & $\sim 1.5$ \\
Alpha & 16 & 3 \\ 
CNO extended & 18 & 3 \\
Nova & 134 & 7 \\
150-isotope & 150 & 7.5 \\
365-isotope & 365 & $\sim 20$ \\
\br       
\end{tabular}
\end{indented}
\end{table}
We will then make a simple estimate of the relative speed of explicit versus
implicit algorithms by multiplying the factor $F$ by the ratio of integration
timesteps for implicit and explicit integrations for a given problem. This
probably underestimates the relative speed of an optimized explicit versus
optimized implicit code for reasons discussed in Ref.\ \cite{guidJCP}, but it
will allow us to place a lower bound on how fast the explicit calculation can
be.

\section{\label{sh:ssCompare} Comparison of QSS Methods with Asymptotic and
Implicit Methods}

In earlier applications of asymptotic and steady-state methods in chemical
reaction networks, evidence was presented that Quasi-Steady-State (QSS)
approximations gave more accurate, stable, and faster solutions than asymptotic
approximations \cite{mott99,mott00}, but that both approximations failed when
applied to the extremely stiff systems characteristic of astrophysical
thermonuclear networks \cite{oran05,mott99}. In a preceding paper we have
investigated the use of explicit asymptotic approximations for extremely stiff
astrophysical networks and concluded that the asymptotic approximation in fact
works quite well for even the stiffest networks, provided that they are not too
close to equilibrium \cite{guidAsy}.  We now wish to revisit the utility of QSS
methods for extremely stiff networks, comparing them with results from both
asymptotic and implicit calculations for some representative extremely stiff
astrophysical networks of varying sizes. In the general case, we shall find that
both asymptotic and QSS methods are capable of solving extremely stiff networks
stably and accurately, but that QSS approximations often allow somewhat larger
timesteps than the corresponding asymptotic approximation calculation. We shall
find that these timesteps for both QSS and asymptotic approximations are often
quite competitive with those of a standard implicit code for systems that are
not near microscopic equilibrium.

\subsection{Appropriate Astrophysical Variables}

For the astrophysical examples given in the remainder of this paper, the generic
population variables $y_i$  (assumed to be proportional to the number density
for the species $i$) will be replaced with the mass fractions $X_i$. These
satisfy
\begin{equation}
X_i  = \frac{n_iA_i}{\rho N\tsub A}
\qquad
\sum_i X_i =1 ,
\label{5.35}
\end{equation}
where $N\tsub A$ is Avogadro's number, $\rho$ is the total mass density, 
$A_i$ is the atomic mass number, and $n_i$ is the number density for the species
$i$.

\subsection{CNO Cycle and the pp-Chains}

\fig{cnoXnetAsyQSScompare} illustrates a comparison of QSS, asymptotic, and
implicit methods for the main branch of the astrophysical CNO cycle,%
\singlefig
{cnoXnetAsyQSScompare}
{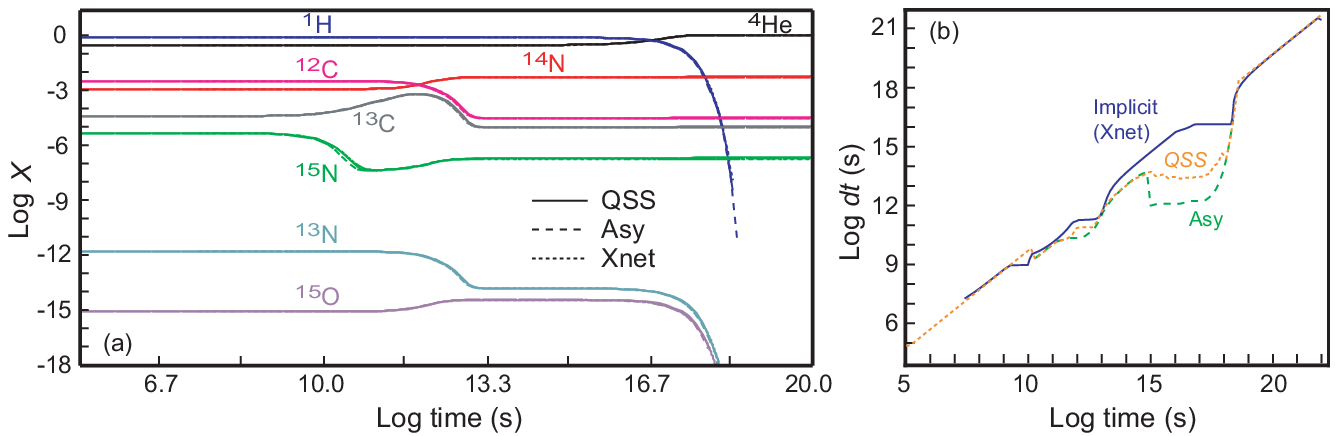}
{0pt}
{\figdn}
{1.0}
{Comparison of asymptotic, quasi-steady-state, and implicit approximations for
the main branch of the CNO cycle at a constant temperature of 20 million K ($T_9
= 0.020$) and
constant density $\rho = 20 \units{g\,cm}^{-3}$, with initial abundances of
solar composition. This network corresponds to 8 isotopes and 18 reaction
couplings, with reaction rates taken from the REACLIB library \cite{raus2000}.
(a)~Isotopic mass fractions. Solid curves are implicit (Xnet \cite{raphcode}),
dashed are QSS and asymptotic.  (b)~Integration timesteps.  The calculation
shown used three iterations for the QSS method, which provided marginal
improvement over a single iteration. }
which is illustrated in Fig.~\ref{fig:cnoCycle}. 
\putfig
{cnoCycle}
{0pt}
{\figdn}
{0.90}
{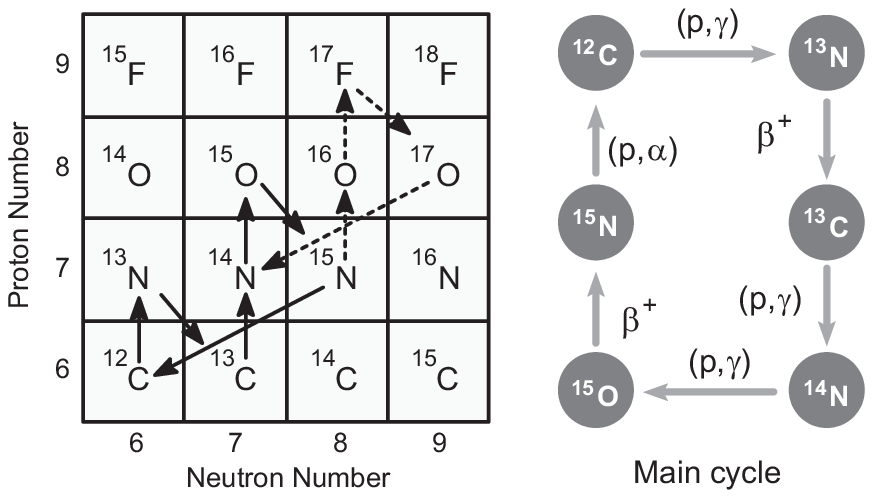}
{The CNO cycle. The main branch of the cycle is illustrated
with solid arrows and a side branch is illustrated with dashed arrows. The main
branch of the CNO cycle is illustrated schematically with more detail on the
right side.}
The calculated mass fractions are almost identical for the three approaches. The
timestepping for the QSS and asymptotic integrations is also very similar,
except for a small region approaching hydrogen depletion ($\log t \sim 17$)
where the QSS method is able to take timesteps 1-2 orders of magnitude larger
than the asymptotic method. This translates into an overall improvement of
roughly a factor of two in time to complete the calculation. The timestepping
for the implicit method is seen to be very similar to that of the two explicit
methods except for the range $\log t = 15$ to $\log t = 18$, where the implicit
method averages 10--100 times larger timesteps than the QSS method. The fastest
stable timestep for a purely explicit method in this calculation is of order 100
seconds and therefore is far off the bottom of the scale in
\fig{cnoXnetAsyQSScompare}. We note that at the end of the calculation the QSS
and asymptotic timesteps are about $10^{20}$ times larger than would be stable
for a purely explicit calculation.

Although this larger timestepping for the implicit method is confined to a small
region in \fig{cnoXnetAsyQSScompare}, this is quite significant for the overall
integration time (a fact partially obscured by the log--log plot). The two
explicit methods spend the bulk (97\% for the asymptotic calculation, for
example) of their total integration times in the region from $\log t = 15$ to
$\log t = 18$, where the implicit integration is taking timesteps 10--100 times
larger than the explicit methods. This translates into 292 total integration
steps for the implicit code, 15,484 for QSS, and 210,398 for the asymptotic
calculation.  The explicit methods, once optimized, may be expected to compute
these timesteps faster, but for a network this small that advantage will likely
be only a factor of two or so (Table \ref{tb:explicitSpeedup}). Thus the QSS
method needs at least another factor of 25 in speed for the calculation of
\fig{cnoXnetAsyQSScompare} to compete with the fastest implicit integration
of the CNO cycle. That is not very important practically for a single
integration of this simple network, since any of the three methods can integrate
it to hydrogen depletion in a fraction of a second on a modern processor, but if
the network were integrated many times the difference would become significant.

In \fig{cnoExtendedQSSAsyXnetXdt} we compare QSS, asymptotic, and fully implicit
calculations for an extended CNO network
\singlefig
{cnoExtendedQSSAsyXnetXdt}
{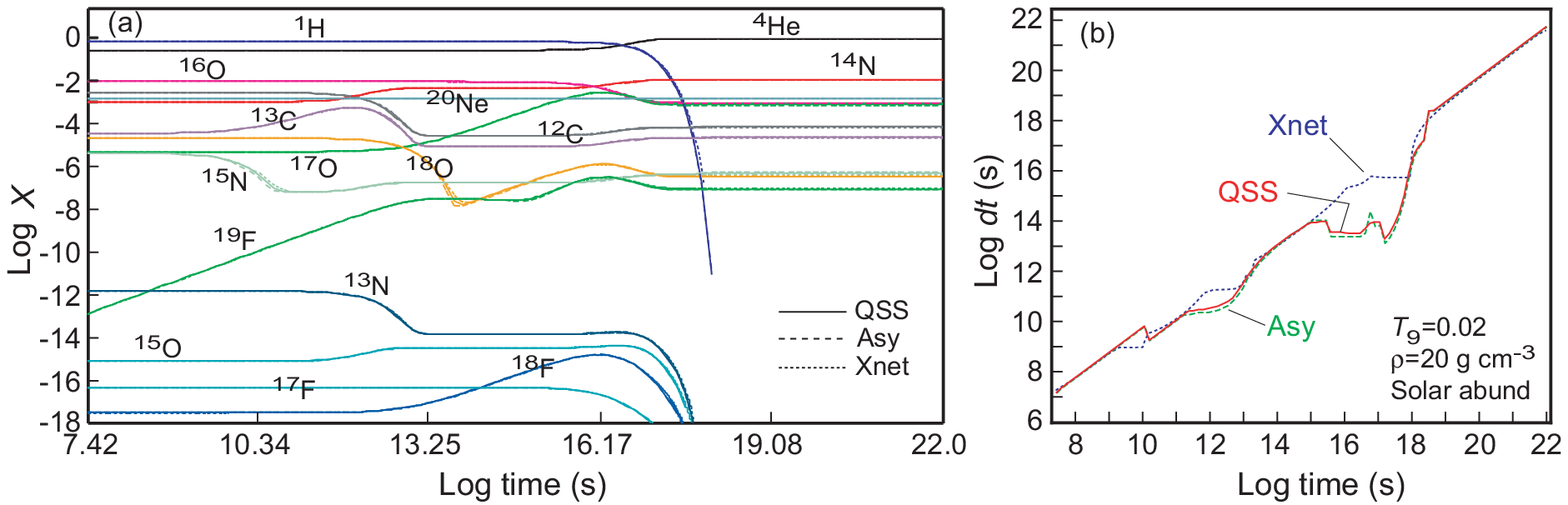}
{0pt}
{\figdn}
{0.73}
{Comparison of implicit (Xnet), quasi-steady-state, and asymptotic
approximations for an extended CNO cycle at a constant temperature of $T_9 =
0.020$ and constant density $\rho = 20 \units{g\,cm}^{-3}$, with initial
abundances of solar composition. This network corresponds to 18 isotopes and
131 reaction couplings, with reaction rates taken from the REACLIB library
\cite{raus2000}. (a)~Isotopic mass fractions. (b)~Integration
timesteps.  The calculation shown used 1 iteration for the QSS method.
 }
(corresponding to the full network shown on the left side of \fig{cnoCycle} plus
several additional isotopes). In this case we see that all three methods give
essentially the same mass fractions and similar timestepping, except for a short
period near $\log t \sim 15-18$ where the implicit calculation takes timesteps
as much as 100 times larger than the other methods. Note that in this example
there is almost no difference between the QSS and asymptotic timestepping,
unlike the case in \fig{cnoXnetAsyQSScompare} where the QSS calculation is
faster. Once again, the log--log scale somewhat obscures that the explicit
methods  need another factor of $\sim$15 in speed to be as fast as the implicit
calculation (the implicit code took 348 total steps, the QSS code took 10,101
steps, and the asymptotic code took 13,095 steps for this case), but all three
methods can compute the network to hydrogen depletion in less than a second of
processor time. As for \fig{cnoXnetAsyQSScompare}, the fastest stable timestep
for a purely explicit method in this calculation is of order 100 seconds and
therefore is off the bottom of the scale in \fig{cnoExtendedQSSAsyXnetXdt}.  At
the end of the calculation the QSS and asymptotic timesteps are about $10^{19}$
times larger than would be stable for a purely explicit calculation.

We speculate that the reason the QSS and asymptotic timesteps lag behind the
implicit method timesteps only for the range $\log t \sim 15-18$  is that this
is roughly the time period when the CNO cycle is running in steady state
(approximately constant abundances for the carbon--nitrogen--oxygen isotopes, as
hydrogen is being converted to helium at a nearly constant rate),  up until the
hydrogen begins to be significantly depleted; see \fig{cnoXnetAsyQSScompare}(a).
In that period the CNO cycle running in steady state establishes a new timescale
in the system, which is the time characteristic of restoring the cycle
equilibrium if it were disturbed. From \fig{cnoXnetAsyQSScompare}(a) we may
estimate that this timescale is approximately $\tau \sim 10^{13}-10^{14}$ s,
since this was the time to establish steady state initially. Neither the
asymptotic nor QSS approximations remove the specialized stiffness associated
with this cycling timescale completely (nor would the partial equilibrium
approximation as we have formulated it, since the cycle does not have reversible
reactions).  Thus the explicit timestep stops growing around $dt \sim
10^{13}-10^{14}$ s because substantially larger explicit timesteps would not be
able to resolve and respond to fluctuations in the CNO equilibrium.  This
remains true until the onset of hydrogen depletion removes this timescale and
the explicit method is again able to increase its timesteps rapidly. This
suggests that a modification of the explicit methods to replace the cycle with
an analytical approximation when it is running near steady state should permit
the explicit methods to increase their timesteps competitively in the time
period $\log t \sim 15-18$.

In the preceding CNO-cycle calculations the implicit method is superior,
performing the integration more than an order of magnitude faster than the
explicit methods.  However, the remarkable result is not that the implicit
algorithm is faster.  Rather, it is that the QSS method has made up almost 19 of
the 20 orders of magnitude difference between the integration speed of a purely
explicit method relative to the implicit method, and that the remaining order of
magnitude is likely because of a highly-specialized stiffness associated with
cycling that has not yet been dealt with in the explicit networks. This
interpretation is bolstered by applying the QSS method to the astrophysical
pp-chains, which are comparable in stiffness  to the CNO cycle but do not
exhibit cycling. Figure \ref{fig:ppChains}%
 \putfig
     {ppChains}
     {0pt}
     {\figdn}
     {0.97}
     {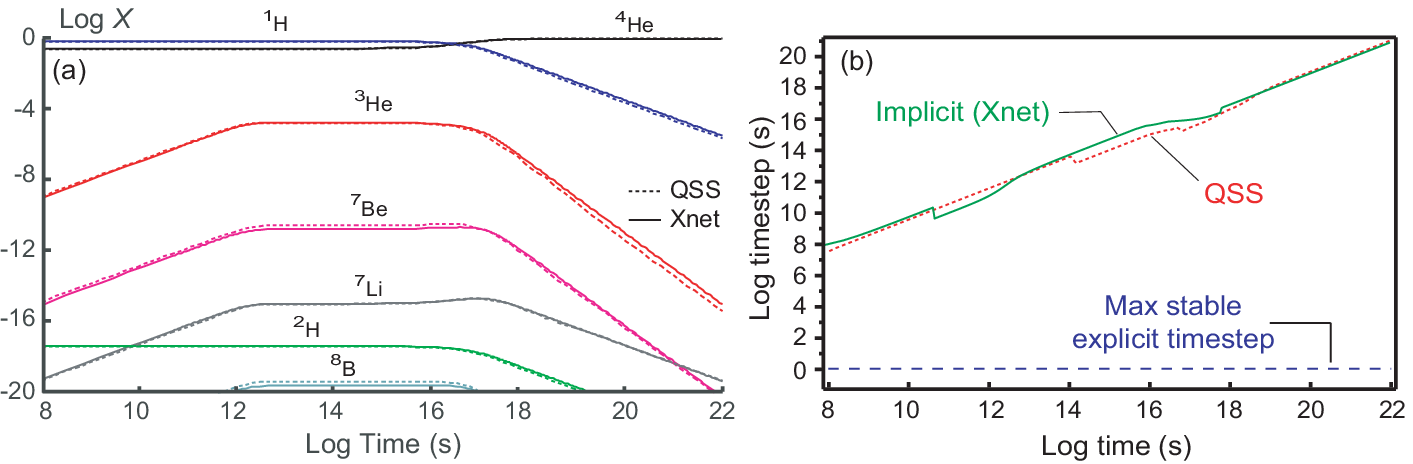}    
{Integration of the pp-chains at constant temperature $T_9 = 0.016$ (where $T_9$
denotes temperature in units of $10^9$ K) and constant density $160
\units{g\,cm}^{-3}$, assuming solar initial abundances. Reaction rates were
taken from the REACLIB library \cite{raus2000}.  (a)~Mass fractions for the
asymptotic method, the QSS method,  and for the standard implicit code Xnet
\cite{raphcode}. (b)~Integration timesteps.}
displays integration of the pp-chains at a constant temperature and density
characteristic of the core in the present Sun using the QSS method and the
implicit backward-Euler code Xnet \cite{raphcode}. In this example we see that
the QSS method has made up essentially all of the more than 20 orders of
magnitude difference between implicit and purely explicit timestepping.  This
gives integration speeds about the same as for the implicit method: the
implicit code required only 176 integration steps versus 286 for the QSS method,
but from Table \ref{tb:explicitSpeedup} each timestep for the 7-isotope pp-chain
network can probably be calculated $\sim$1.5 times faster using the explicit
code.

\subsection{\label{sn1aDetonation}Type Ia Supernova Detonation Waves}

In \fig{Xdt_T9_10rho5e7Alpha}%
\singlefig
{Xdt_T9_10rho5e7Alpha}
{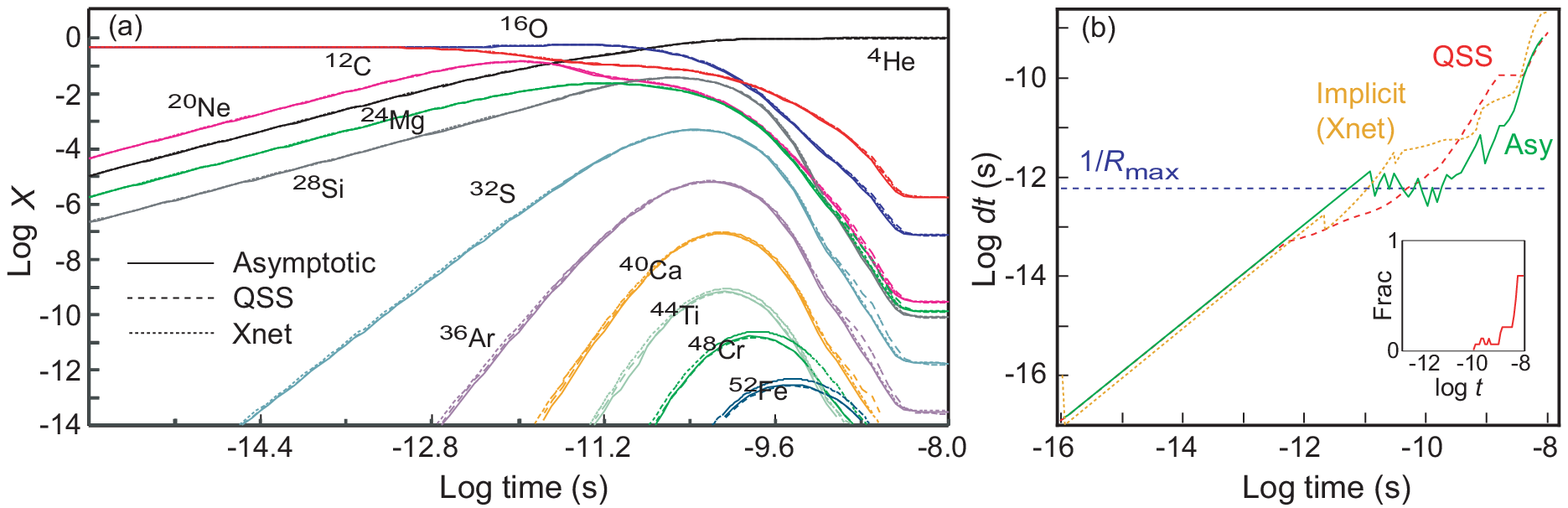}
{0pt}
{\figdn}
{0.72}
{Comparison of asymptotic and quasi-steady-state approximations for an alpha
network at a constant temperature of $T_9 = 10$ and constant density $\rho =
5\times 10^7 \units{g\,cm}^{-3}$, with initial equal mass fractions of
\isotope{12}{C} and \isotope{16}{O}, and reaction rates specified by REACLIB
\cite{raus2000}. The network contained 16 isotopes with 46 reaction couplings.
Also shown are results from the implicit code Xnet \cite{raphcode}. (a)~Isotopic
mass fractions. (b)~Integration timesteps. Solid curves are asymptotic,
dashed curves are QSS, and dotted curves are implicit; the dotted blue curve
estimates the maximum stable purely explicit timestep. The inset to (b) shows
the fraction of reactions equilibrated as a function of time. The calculation
shown used one iteration for the QSS method; additional iterations did not lead
to substantial improvement. }
we compare asymptotic and QSS calculations for an alpha-particle network at a
constant temperature of $T_9 = 10$ and constant density $\rho = 5\times 10^7
\units{g\,cm}^{-3}$, which represents conditions that might be found for a
strong detonation wave in a Type Ia supernova simulation.  We see that the mass
fractions computed in the two cases are essentially the same, except for some
small differences in the weaker populations near $\log t \sim -9$.  At earlier
times the asymptotic method gives somewhat larger timesteps but at intermediate
times corresponding to maximal burning the QSS timesteps are as much as an order
of magnitude larger.  The QSS and asymptotic integration times are also seen to
be rather competitive with those of the implicit calculation. The total
calculation required 1464 asymptotic timesteps, 714 QSS timesteps, and 329
implicit timesteps. Since an explicit timestep can be computed about 3 times
faster by the explicit methods relative to the implicit method for this
16-isotope network (Table \ref{tb:explicitSpeedup}), equivalently-optimized
versions of all three methods would be rather similar in speed. 

Consulting the inset to \fig{Xdt_T9_10rho5e7Alpha}(b), we see that almost no
reactions become microscopically equilibrated until very late in the
calculation, which explains the competitive explicit QSS and asymptotic
timesteps over most of the integration range.  Although the amount of partial
equilibrium is small until late in the preceding calculation, this still has a
significant negative impact on the QSS and asymptotic timesteps. In Ref.\
\cite{guidPE} we shall implement a partial equilibrium formalism to deal with
this.  If those methods are applied to the present problem, the required number
of integration steps is reduced from 714 to 313 for the QSS method and from 1464
to 322 for the asymptotic method. Thus, with partial equilibrium accounted for
the calculation of \fig{Xdt_T9_10rho5e7Alpha} would become several times faster
for both QSS and asymptotic methods relative to the implicit calculation,
by virtue of the speedup factor of about 3 for the explicit method from Table
\ref{tb:explicitSpeedup}.

The QSS method can be iterated  to improve the solution \cite{mott00}.  The
calculation shown in \fig{cnoXnetAsyQSScompare} used three QSS iterations, but a
single iteration gave results almost as good.  The calculation shown in
\fig{Xdt_T9_10rho5e7Alpha} used a single iteration and was not significantly
improved by additional iterations. In our tests on very stiff thermonuclear
networks, we have found that iterating the QSS solution does not generally give
significantly better results than  a single-iteration calculation, but can
improve the speed corresponding to a given precision in some cases by factors of
two.

In \fig{X_dt365T9_10rho_5e7QSS}
\singlefig
{X_dt365T9_10rho_5e7QSS}
{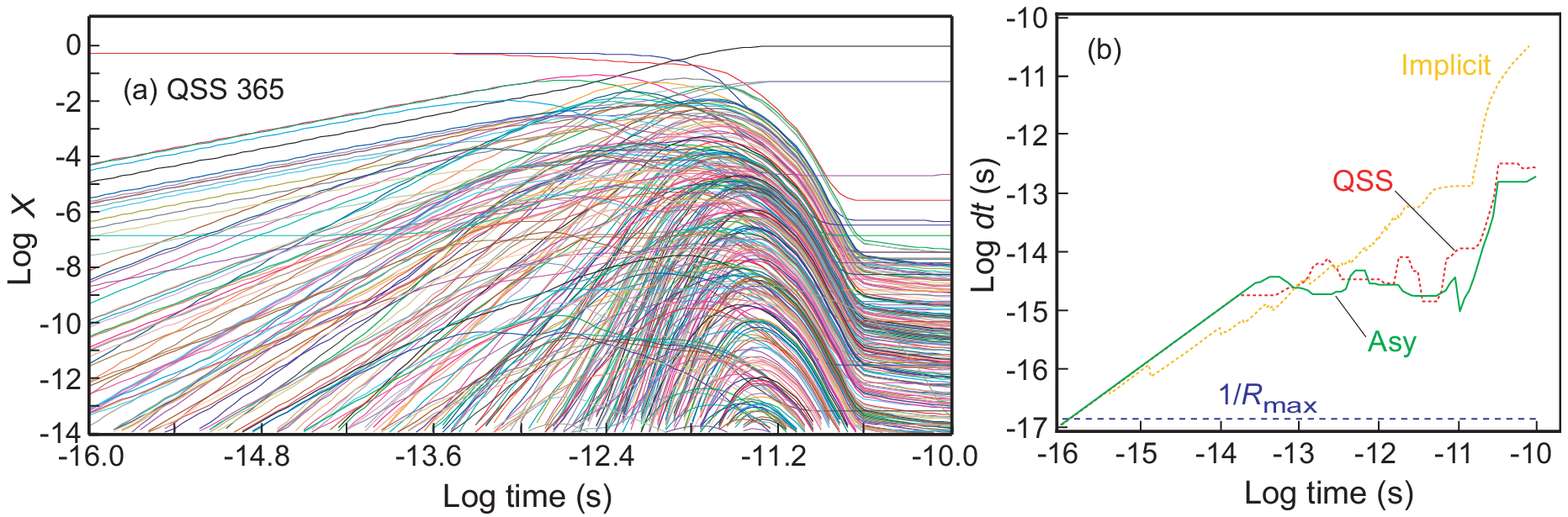}
{0pt}
{\figdn}
{0.70}
{A 365-isotope network integrated at constant temperature $T_9 = 10$ and
constant density $\rho = 5\times 10^7 \units{g\,cm}^{-3}$, for an initial
concentration of half \isotope{12}{C} and half \isotope{16}{O} by mass fraction.
The network contained 365 isotopes and 4325 reaction couplings, with the
reaction rates supplied by  REACLIB \cite{raus2000}.(a)~Isotopic mass fractions
for the quasi-steady-state (QSS) method. (b)~Integration timesteps for the
asymptotic method (solid green curve),  the QSS method (dashed red curve),  and
the implicit code Xnet \cite{raphcode} (dotted orange curve). The maximum stable
timestep for a normal explicit calculation (dashed blue curve) was estimated as
the inverse of the fastest rate in the network. }
we display mass fractions for the QSS method and compare timestepping for the
QSS, asymptotic, and explicit methods for the same conditions as in
\fig{Xdt_T9_10rho5e7Alpha}, but for a 365-isotope network. Since there are so
many mass-fraction curves in \fig{X_dt365T9_10rho_5e7QSS}(a), we do not attempt
to compare them directly with an asymptotic or implicit calculation. 
However, in Ref.\ \cite{guidAsy} we established the equivalence of mass
fractions calculated by standard implicit and asymptotic approximations, and 
in \fig{compareAsyQSSdE_T9_10rho5e-7_365}(a)
\singlefig
{compareAsyQSSdE_T9_10rho5e-7_365}
{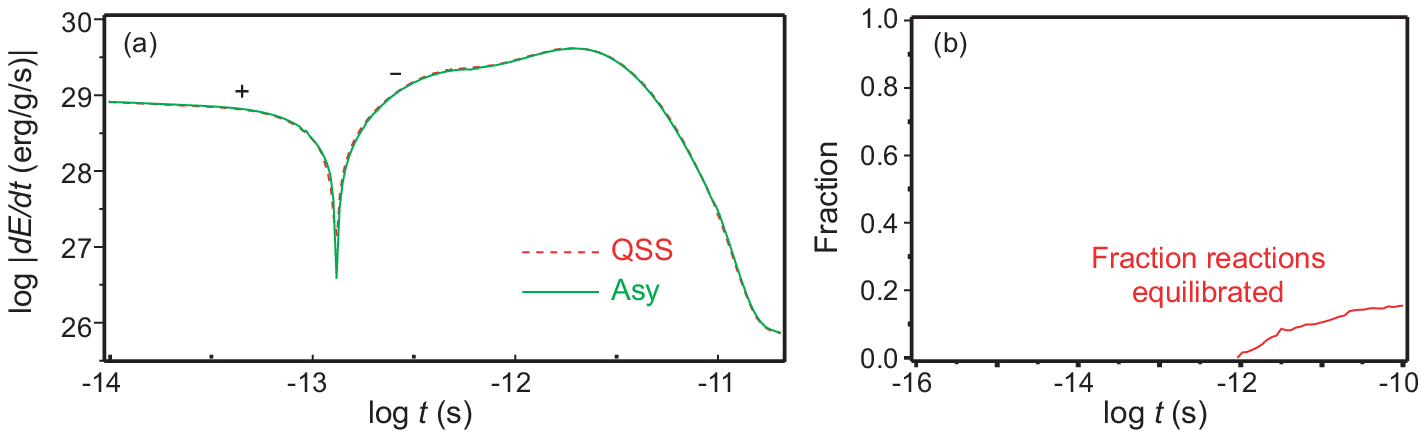}
{0pt}
{\figdn}
{0.93}
{(a)~Comparison of differential energy production calculated by QSS (dotted red
curve) and asymptotic approximations \cite{guidAsy} (solid green curve) for the
365-isotope network of \fig{X_dt365T9_10rho_5e7QSS}. The energy production
changes sign near $\log t = -13$, so we plot the log of the absolute value of
$dE/dt$ and indicate its sign on the curve. (b)~Fraction of reactions
microscopically equilibrated in the QSS calculation.}
we compare the differential energy production (a strong proxy for evolution of
the isotopic number densities) for the network in \fig{X_dt365T9_10rho_5e7QSS}
calculated by QSS and asymptotic methods. The curves are in almost perfect
agreement, and the integrated energy release corresponding to the simulation of
\fig{X_dt365T9_10rho_5e7QSS} differed by less than 0.2\% between QSS and
asymptotic calculations.

For this example the QSS calculation used two iterations of the
predictor--corrector algorithm \eqnoeq{qss1.2}, which permitted almost a factor
of two larger timestep size than for one iteration. Computing the rates is the
most time-consuming operation in an explicit timestep. Since a
predictor--corrector iteration recomputes the fluxes by multiplying the rates by
the new populations from the predictor step but does not recompute the rates, it
does not cost much. In this case, once the rates have been calculated at the
current temperature and density, each iteration increases the time to compute
the timestep by only a few percent.

The methods all take similar timesteps until $\log t \sim -12$,
after which the QSS calculation takes somewhat larger timesteps than the
asymptotic method, while the implicit calculation takes timesteps that average
about 10 times larger than the QSS method. As a result, for the entire
calculation the implicit code required 444 integration steps, the QSS code
required 4398 steps, and the asymptotic code required 9739 steps. The timestep
advantage of the implicit code over the QSS code by about a factor of 10 will be
approximately canceled by the $\sim 20$ times faster computation of each
timestep by the explicit code for a 365-isotope network (see Table
\ref{tb:explicitSpeedup}). Thus, for the case in \fig{X_dt365T9_10rho_5e7QSS} we
expect that for optimized codes the QSS and implicit methods would have similar
speeds, and the asymptotic method would be about a factor of two slower.

In \fig{compareAsyQSSdE_T9_10rho5e-7_365}(b) we plot the fraction of reactions
in the network that become microscopically equilibrated.  The reason that the
implicit code is able to take larger timesteps than the explicit codes for $\log
t > -12$ in the 365-isotope case now becomes clear: that is exactly where
partial equilibrium begins to play a role.  Because the fraction of
partially-equilibrated reactions  reaches only $\sim$15\% in this calculation,
the explicit methods are still able to compete favorably, but as we have already
seen for the alpha network of \fig{Xdt_T9_10rho5e7Alpha}, and as we shall see
further later in this paper and in Ref.\ \cite{guidPE}, even a small
partial-equilibrium fraction can have a large negative influence on the explicit
integration timestep.   The impact on the total integration time for the present
examples will be amplified because the QSS calculation in
\fig{X_dt365T9_10rho_5e7QSS} expends more than 90\% of its integration steps for
times where partial equilibrium is significant (while in the alpha network of
\fig{Xdt_T9_10rho5e7Alpha} the corresponding fraction is about 75\%). Thus, we
may expect that a proper treatment of partial equilibrium in this case should
lead to an explicit QSS or asymptotic timestep that is much larger, implying a
substantial speed advantage for each of the two explicit methods versus the
implicit method,  once proper account has been taken of partial equilibrium.

\subsection{\label{tidalSupernova} Tidal Supernova Simulation}

A comparison of asymptotic and QSS mass fractions and timesteps is shown in
\fig{tidalAlphaAsyQSS_X_dt}%
\singlefig
{tidalAlphaAsyQSS_X_dt}
{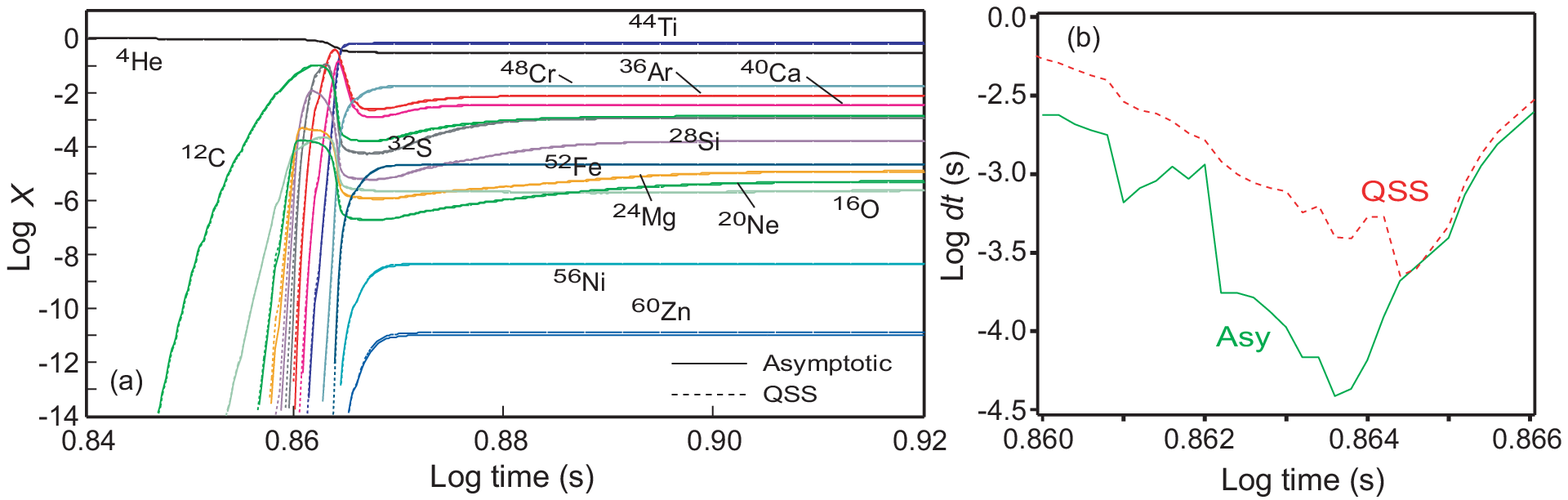}
{0pt}
{\figdn}
{0.70}
{Comparison of asymptotic and quasi-steady-state approximation calculations for
an alpha network calculation using the hydrodynamical profile shown in
\fig{tidalAlphaHydroProfile}. (a)~Mass fractions (solid lines asymptotic, dashed
lines QSS; they are almost indistinguishable).  (b)~Integration timesteps for
asymptotic (solid) and QSS (dashed) in the strong burning region. The QSS
calculation shown used three iterations but one or two iterations gave similar
results. }
for an alpha network with a hydrodynamical profile characteristic of a supernova
induced by tidal interactions in a white dwarf (illustrated in
\fig{tidalAlphaHydroProfile}).%
\singlefig
     {tidalAlphaHydroProfile}
     {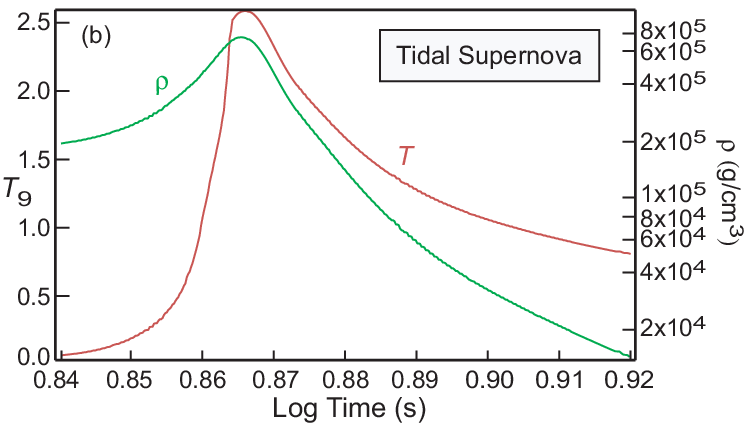}
     {0pt}
     {\figdn}
     {1.0}
{Hydrodynamical profiles for temperature and density under tidal supernova
conditions \cite{tidalSupernova}.}
% %
% %
We see that in the critical strong-burning region the QSS approximation is able
to take timesteps that are about an order of magnitude faster than the
asymptotic method. (Outside this region the timesteps are similar for the two
methods.) The timestepping  over the entire integration range is compared for
QSS, an asymptotic calculation, and the implicit code Xnet in
\fig{tidalAlpha_dtPlusFractionQSS}(a).
\singlefig
     {tidalAlpha_dtPlusFractionQSS}
     {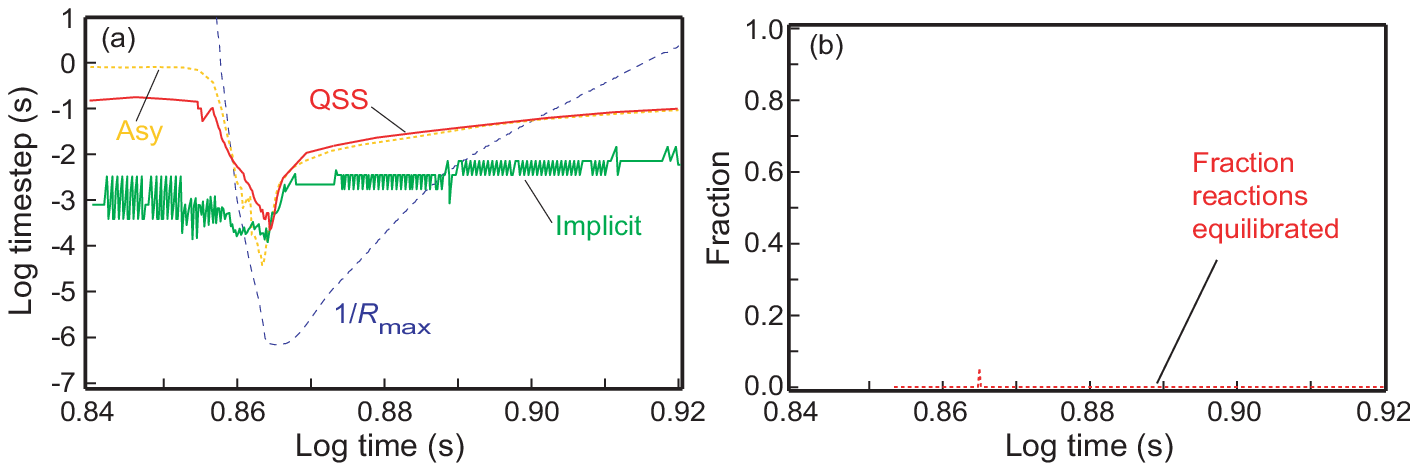}
     {0pt}
     {\figdn}
     {0.90}
{(a)~QSS integration timesteps (solid red), Asymptotic integration timesteps
(dotted orange), integration steps for the implicit code Xnet \cite{raphcode}
(solid green), and maximum stable purely-explicit step (dashed blue), for the
calculation in \fig{tidalAlphaAsyQSS_X_dt}. (b)~Fraction of isotopes that become
asymptotic and fraction of reactions equilibrated in the network.}
% %
% %
The QSS timestepping (242 total integration steps) is somewhat better than for
the asymptotic method (480 steps) and considerably better than for the implicit
code (2136 steps).  For an alpha network an
optimized explicit code can compute timesteps about three times as fast as an
implicit code (Table \ref{tb:explicitSpeedup}), so we may estimate that the QSS
code is capable of calculating this network some 15 times faster, and the
asymptotic code  perhaps 10 times faster, than the implicit code. Results almost
as good as those presented above for networks under tidal supernova conditions
using the QSS method
have been found in Refs.~\cite{feg11a,feg11b} using the explicit asymptotic
method. Although a different set of reaction network rates was used in these
references, the explicit asymptotic method was  found to be highly
competitive with standard implicit methods for the tidal supernova problem.
 
The good QSS and asymptotic timestepping for this case is primarily because
essentially no reactions in the network come into equilibrium, as illustrated in
\fig{tidalAlpha_dtPlusFractionQSS}(b). Note in this connection that the flat
mass fraction curves at late times in \fig{tidalAlphaAsyQSS_X_dt}(a) are not a
result of equilibrium, but rather of reaction freezeout caused by the
temperature and density dropping quickly at late times as the system expands
(see \fig{tidalAlphaHydroProfile}).  It is this rapid decrease of all thermal
reaction rates to zero at later times that prevents microscopic equilibration
from playing a significant role for this case. 

The dashed blue curve in \fig{tidalAlpha_dtPlusFractionQSS}(a) represents the
estimated fastest stable purely-explicit timestep.  By comparing this curve with
the QSS and asymptotic timestep curves we see that, unlike most of the cases we
are investigating, this system is only moderately stiff.  The maximum difference
between the actual timestep and the maximum stable purely-explicit timestep is
about 4 orders of magnitude.  Furthermore, for $\log t < 0.86$ and $\log t >
0.9$ the maximum stable purely-explicit timestep is {\em larger} than the actual
timestep; thus in these regions the stiffness instability plays no role in
setting the explicit QSS or explicit asymptotic timestep. This might suggest
that it is the relatively moderate stiffness of this problem compared with most
in astrophysics that makes the QSS and asymptotic methods particularly efficient
for this case. However, comparison with a number of other cases (see for
example, the nova calculation in \S\ref{novaExplosions}) indicates that this is
not correct:  it is not the {\em amount} of stiffness (as inferred from
differences in ranges of timescales in the problem) but rather the {\em nature}
of the stiffness that is crucial. The QSS and asymptotic methods are capable of
removing many orders of magnitude of stiffness caused by macroscopic
equilibration (for example, see \fig{nova125D_dtPlusfractionQSS}(a) below), but
are poor at removing stiffness caused by microscopic equilibration.

\subsection{\label{novaExplosions} Nova Explosions}

Let us turn now to an example involving a large and extremely stiff network. In
\fig{nova125D_XplusHydroProfileQSS}(a)%
\singlefig
     {nova125D_XplusHydroProfileQSS}
     {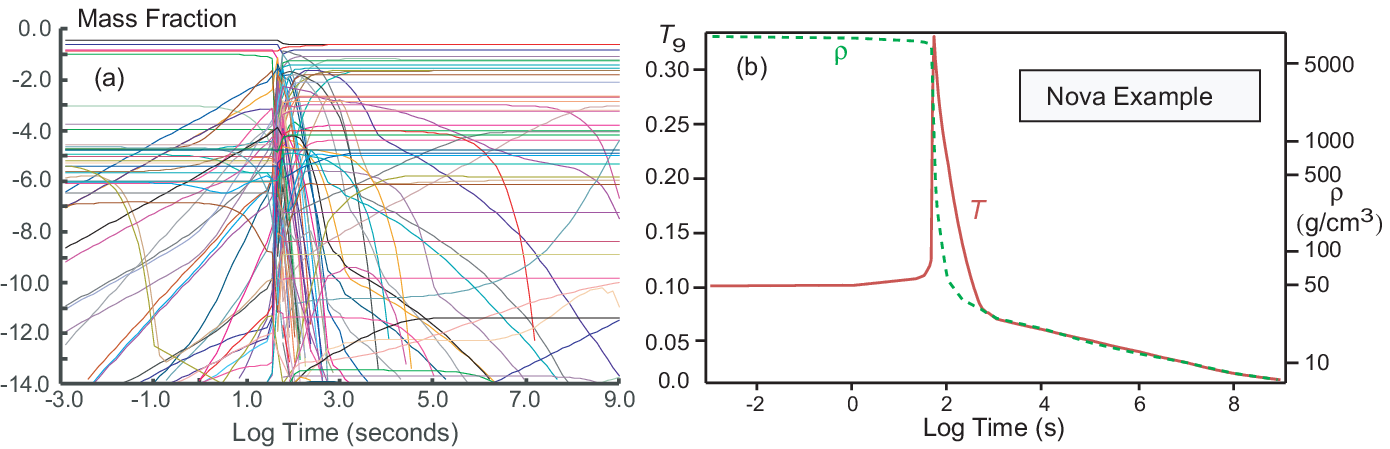}
     {0pt}
     {\figdn}
     {1.0}
{(a)~Mass fractions for a network under nova conditions, corresponding to the
hydrodynamical profile shown in (b). The calculation used the QSS method and a
network containing 134 isotopes coupled by 1531 reactions, with rates taken from
the REACLIB library \cite{raus2000} and initial abundances enriched in heavy
elements \cite{parete-koon03}.}
% %
% %
we illustrate a calculation using the explicit QSS algorithm with a
hydrodynamical profile displayed in \fig{nova125D_XplusHydroProfileQSS}(b) that
is characteristic of a nova outburst. Given the large number of mass-fraction
curves, we do not attempt to compare them directly with an asymptotic or
implicit calculation, but we note that the total integrated energy release 
corresponding to the simulation of \fig{nova125D_XplusHydroProfileQSS} was
within 1\% of that found for the same network using the  explicit asymptotic
approximation in Ref.\ \cite{guidAsy}. The integration timesteps for the
calculation in \fig{nova125D_XplusHydroProfileQSS}(a) are displayed in
\fig{nova125D_dtPlusfractionQSS}(a).%
\singlefig
     {nova125D_dtPlusfractionQSS}
     {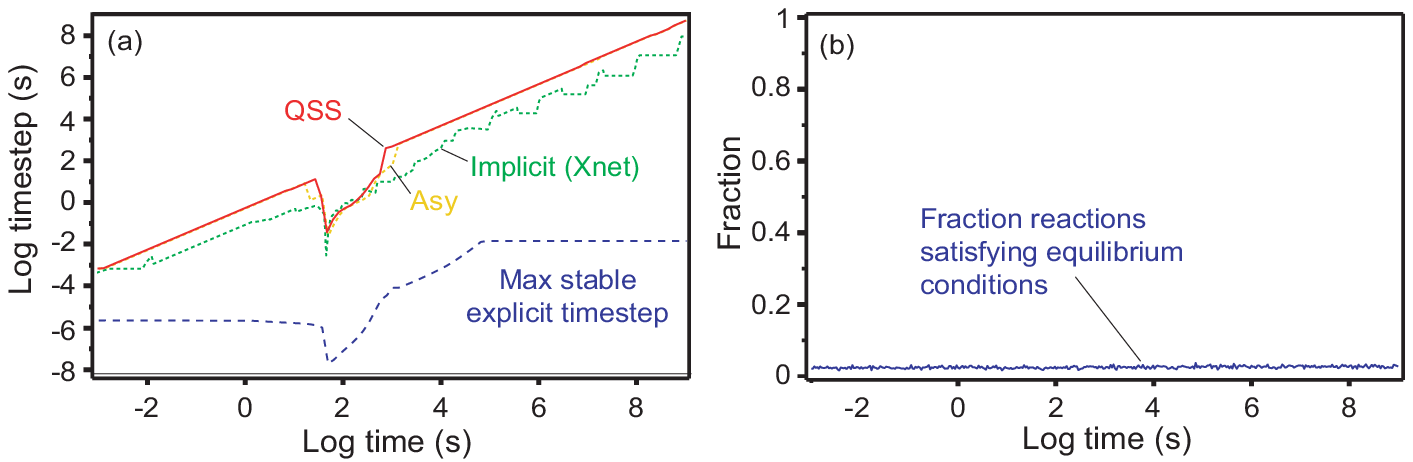}
     {0pt}
     {\figdn}
     {0.95}
{(a)~Timesteps for integration of \fig{nova125D_XplusHydroProfileQSS}. The solid
red curve is from the QSS calculation.  The dotted orange curve is from an
explicit asymptotic calculation \cite{guidAsy}. The dotted green curve is from
an implicit integration using the backward-Euler code Xnet \cite{raphcode}.  The
dashed blue curve estimates the largest stable fully explicit timestep as the
inverse of the fastest rate in the system. (b)~Fraction  of reactions that reach
partial equilibrium in the QSS calculation.}
% %
% %
Once burning commences, the QSS solver (solid red curve in
\fig{nova125D_dtPlusfractionQSS}(a)) takes timesteps that are from $10^6$ to
$10^{10}$ times larger than would be stable for a normal explicit integration. 

The explicit QSS timesteps  illustrated in \fig{nova125D_dtPlusfractionQSS}(a)
are somewhat larger than those of our asymptotic solver (dotted orange curve),
and comparable to or greater than those for a typical implicit code over the
whole integration range, as may be seen by comparing with the implicit (backward
Euler) calculation timestepping curve shown in dotted green. In this calculation
the implicit method required 1332 integration steps, the explicit asymptotic
calculation required 935 steps, and the QSS method required 777 steps. Given
that for a network with 134 isotopes the explicit codes should be able to
calculate an integration timestep about 7 times faster than an implicit code
because they avoid the manipulation of large matrices (Table
\ref{tb:explicitSpeedup}), these results suggest that the explicit QSS code is
capable of calculating the nova network more than 10 times faster and  the
explicit asymptotic code  more than 5 times faster than a state-of-the-art
implicit code. 

This impressive integration speed for both the QSS and asymptotic methods
applied to a large, extremely stiff network is possible because few reactions
reach microscopic equilibrium during the simulation, as illustrated in
\fig{nova125D_dtPlusfractionQSS}(b). Thus the entire nova simulation, just as
for the tidal supernova simulation and the Type Ia supernova detonation wave
simulation until very late in the calculation, lies within a domain where we
expect both the QSS and asymptotic explicit methods to be highly effective in
removing stiffness from the network.  In Refs.~\cite{feg11a,feg11b} the
explicit asymptotic method was applied to a nova simulation.  Although
this calculation differed from the present one in using asymptotic methods, a
different nova hydrodynamical profile, and a different reaction library, results
rather similar to those presented above were obtained. We conclude that the
explicit QSS and asymptotic methods may intrinsically be considerably faster
than a state-of-the art implicit code for simulations of nova outbursts.

\section{\label{sh:noncompetive} Non-Competitive QSS Timesteps in the
Approach to Equilibrium}

Except for late in the calculation for the supernova detonation wave in
\S\ref{sn1aDetonation}, the examples shown to this point have involved networks
in which few reactions have become microscopically equilibrated by the criteria
of \S\ref{sh:large-stiff}. For such cases we have seen that the integration
speed for QSS and asymptotic explicit methods is often comparable to, and in
some cases exceeds, that for current implicit codes.  Let us now turn to a
representative example where this is no longer true. The calculation in
\fig{compare_dt_asy_qssT9_5rho1e8Alpha}%
\singlefig
{compare_dt_asy_qssT9_5rho1e8Alpha}
{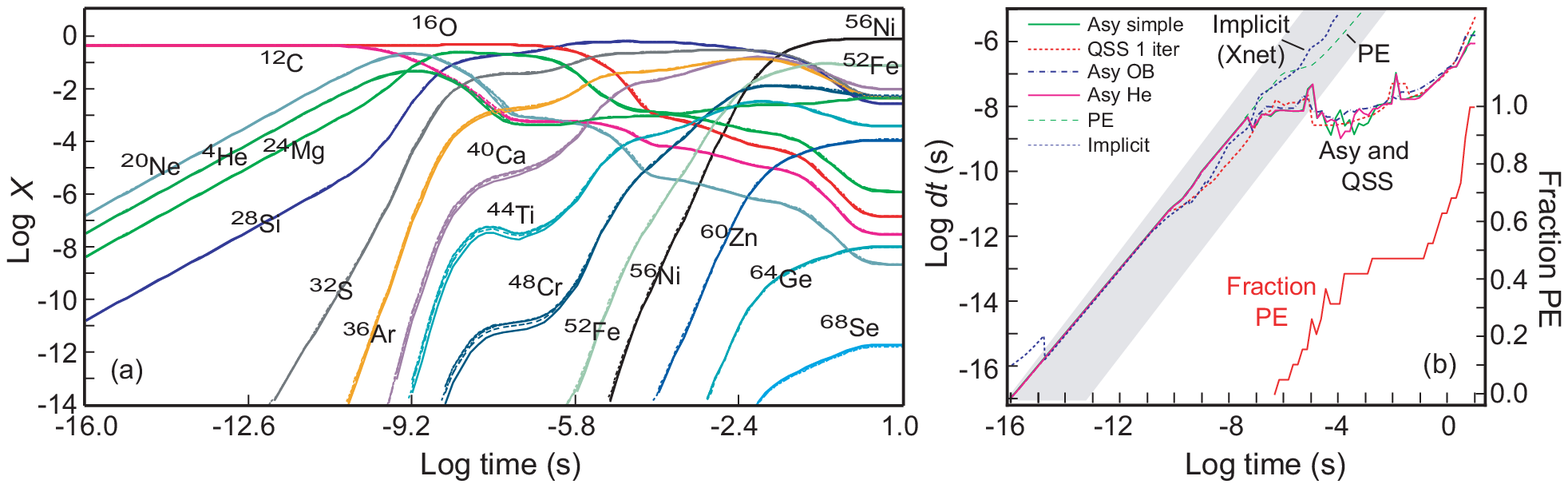}
{0pt}
{\figdn}
{0.70}
{Comparison of asymptotic and quasi-steady-state approximations for an alpha
network with constant temperature $T_9 = 5$ and constant density of $10^8
\units{g\,cm}^{-3}$, using REACLIB rates \cite{raus2000} and initial equal mass
fractions of \isotope{12}{C} and \isotope{16}{O}.   (a)~Isotopic mass fractions.
(b)~Integration timesteps (left axis) and fraction of reaction in partial
equilibrium (right axis). The gray shaded area represents roughly the region
that the explicit timestep profile must lie in to have a chance to compete with
implicit methods. The different asymptotic methods are labeled Asy and are
described in Ref.\ \cite{guidAsy}. The implicit calculation was made with
Xnet \cite{raphcode} and the dashed green line (PE) represents the
timestepping for a partial equilibrium calculation that will be discussed in a
later publication \cite{guidPE}.  The QSS calculation was run with a single
iteration. }
compares QSS and several different asymptotic approximations with an implicit
calculation for an alpha network at a constant temperature and density
characteristic of a Type Ia supernova
explosion. We may draw two important conclusions from these results.

\begin{enumerate}
 \item 
Although there are some differences among the QSS and various asymptotic
methods, we see that they all give essentially the same results, with
timestepping that is rather similar, though timestepping differences of up to
factors of 5-10 may be found in localized time regions. All of the QSS and
asymptotic cases shown have integrated final energies that lie within 1\% of
each other and their total integration times are all within 25\% of each other.
\item
The  QSS method and the various asymptotic methods all give timesteps that
potentially are
competitive with implicit methods  at early
times, but they fall far behind at late times. 
\end{enumerate}
The reason for the non-competitive nature of the asymptotic and QSS timestepping
at late times in this calculation can be seen clearly from the solid red curve
on the right of \fig{compare_dt_asy_qssT9_5rho1e8Alpha}(b), which represents the
fraction of reactions in the network that satisfy partial equilibrium
conditions. We see from this and previous results that generally asymptotic and
quasi-steady-state approximations work very well as long as the network is
well-removed from equilibrium, but as soon as significant numbers of reactions
in the network become microscopically equilibrated the asymptotic and QSS
timestepping begins to fall far behind. In this example, we see that even a 10\%
fraction of equilibrated reactions has a significant negative impact on the
asymptotic and QSS timestepping. 

In earlier sections we have presented evidence that, well-removed from
equilibrium, quasi-steady-state methods can provide stable and accurate
integration of the stiffest large networks with timesteps that are comparable to
those employed in standard implicit and semi-implicit solvers. In
practice, for astrophysical thermonuclear networks this means that timesteps are
typically from $0.1$ to $0.001$ of the current time over most of the integration
range, except for short time periods where very strong fluxes are being produced
and timesteps may need to be shorter to maintain accuracy.  Since explicit
methods can generally compute each timestep substantially faster than for
implicit methods in large networks, this suggests that asymptotic or QSS solvers
 offer a viable alternative to implicit solvers under those conditions. 

However, the preceding statements are no longer true when substantial numbers of
reaction pairs in the network begin to satisfy equilibrium conditions.  Then the
generic behavior for both steady-state and asymptotic approximations is that
exhibited in \fig{compare_dt_asy_qssT9_5rho1e8Alpha}, with the explicit timestep
becoming constant or only slowly increasing with integration time. We shall
explain in the third paper of this series \cite{guidPE} the reason for the loss
of efficiency in asymptotic and steady-state methods as equilibrium is
approached: these approximations remove  major sources of stiffness, but near
(microscopic) equilibrium a fundamentally new kind of stiffness enters the
equations that is not generally removed by either QSS or asymptotic
approximations. Dealing with the stiffness brought on by the approach to
microscopic equilibrium requires that asymptotic or QSS methods be augmented by
a new algebraic approximation tailored specifically to turn equilibrium from a
liability into an asset.  

In the third paper of this series \cite{guidPE} we shall describe a new
implementation of {\em partial equilibrium methods} that can be used in
conjunction with asymptotic or QSS methods to increase the explicit timestepping
by orders of magnitude in the approach to equilibrium. In that paper we will
give examples suggesting that this  partial equilibrium method is capable of
competing strongly with implicit methods across the entire range of interesting
physical integration times for a variety of extremely stiff networks. We give a
preview of those results in \fig{compare_dt_asy_qssT9_5rho1e8Alpha}(b). The
dashed green line labeled PE corresponds to an explicit partial equilibrium plus
asymptotic approximation that is seen to exhibit highly-competitive timestepping
relative to that of the implicit calculation, even as the network approaches
equilibrium.

\section{Conclusions}

In this paper we have compared quasi-steady-state (QSS) calculations with
asymptotic and implicit calculations for extremely stiff networks and concluded
that
\begin{enumerate}
\item
QSS and asymptotic methods give similar results, but QSS timesteps are generally
at least as large as for asymptotic methods, and can be larger by as much as
an order of magnitude in some cases.
 \item 
Both QSS methods and asymptotic methods are uniformly capable of stable,
accurate solutions, even for extremely stiff thermonuclear networks, with
timesteps that are substantially larger than those for standard explicit
methods. The only question then is whether such methods can use large enough
integration timesteps to be competitive with implicit methods.
\item
As for asymptotic methods \cite{guidAsy}, QSS methods give integration speeds
that compete with or even exceed that for implicit methods in extremely stiff
networks as long as the system is well removed from (microscopic) equilibrium,
but fail to
deliver competitive timesteps in the approach to equilibrium. Solution of this
problem will require explicit partial equilibrium methods that we shall discuss
in Ref.\ \cite{guidPE}.
\end{enumerate}
Thus, we find compelling evidence that  quasi-steady-state and asymptotic
methods may have significant application in the integration of large networks
for even the stiffest systems if they are not close to microscopic equilibrium.

Although these conclusions indicate that asymptotic and QSS methods must be
supplemented by partial equilibrium methods to make explicit integration viable
across a full range of stiff problems, the results of this paper and those of
Ref.\ \cite{guidAsy} suggest that the practical utility of the asymptotic and
QSS methods alone for application in astrophysics and many other fields may be
substantial. As we have seen, there are important, extremely stiff problems for
which the system never becomes significantly equilibrated. This is most likely
to occur in explosive scenarios, where we expect rapid expansion on a
hydrodynamical timescale. The expansion will typically lead to reaction
freezeout for those reactions that are strongly temperature-dependent, and in
rapidly-changing environments this may occur before the system has had time to
establish significant microscopic equilibration. The nova calculation of
\S\ref{novaExplosions} and the tidal supernova calculation of
\S\ref{tidalSupernova} are examples of realistic situations where this occurs.
For such problems we have presented evidence that quasi-steady-state or
asymptotic approximations alone (even without partial equilibrium methods) may
provide integration speeds that rival or even substantially exceed those for the
best current implicit methods, particularly for larger networks. 

Even for problems involving large networks coupled to hydrodynamics where the
preceding is not true globally, it will usually be that for many hydrodynamical
zones over various time ranges the conditions will not favor equilibration.
Thus, at each hydrodynamical timestep it may prove most efficient to integrate
the reaction networks for all zones not exhibiting significant reaction-network
equilibration using explicit QSS or asymptotic methods.  For those zones
exhibiting significant equilibration at a given hydrodynamical timestep, more
work will be required to determine whether standard implicit methods such as
backward Euler, or explicit asymptotic or QSS augmented by partial equilibrium
methods are most efficient. If the latter turns out to be true, it likely will
be most useful to integrate all zones with an asymptotic or QSS plus partial
equilibrium method, but it could turn out that the most efficient approach is a
hybrid reaction network algorithm capable of switching among asymptotic plus
partial equilibrium, QSS plus partial equilibrium, and implicit methods as
conditions dictate.

\section{Summary}

Previous examinations of numerical integration for stiff reaction networks have
concluded rather consistently that explicit methods have little chance of
competing with implicit methods for stiff networks because explicit methods are
unable to take large enough stable timesteps. {\em Numerical Recipes}
\cite{press92} goes so far as to state unequivocally that ``For stiff problems
we {\em must} use an implicit method if we want to avoid having tiny
stepsizes.''  These sentiments to the contrary, asymptotic and steady state
approximations have had some success in extending explicit timesteps to usable
sizes for moderately stiff networks, such as those employed for various chemical
kinetics problems \cite{mott99,mott00}. However, such methods were found
previously to be inadequate when applied to the extremely stiff networks
encountered commonly in astrophysical applications, giving very incorrect
results, with timestepping not competitive with implicit and semi-implicit
methods, for thermonuclear networks operating under the extreme conditions of a
Type Ia supernova explosion  (see Ref.\ \cite{oran05} and the discussion in
Ref.\ \cite{mott99}, in particular).

This paper, the preceding one on asymptotic methods \cite{guidAsy}, and the
following one on partial equilibrium methods \cite{guidPE}, reach rather
different conclusions, presenting evidence that algebraically-stabilized
explicit methods work and may be capable of timesteps competitive with those for
implicit methods in a variety of highly-stiff reaction networks. Since explicit
methods  scale linearly and therefore more favorably than implicit
algorithms with network size, our results suggest that algebraically-stabilized
explicit algorithms may be far more competitive than previously thought in a
variety of applications.   Of particular significance is that these new
approaches may permit for the first time the coupling of physically-realistic
kinetic equations to multidimensional fluid dynamics in a variety of
disciplines.

\begin{ack}
We thank Raph Hix, Tony Mezzacappa, Elisha Feger, and Jay Billings for
illuminating discussions, and Jay Billings for a careful reading of the
manuscript. Research was sponsored by the  Office of Nuclear Physics, U.S.
Department of Energy.
\end{ack}

\clearpage

\bibliographystyle{unsrt}

\end{document}